# Quantifying the Consequences of Catheter Steerability Limitations on Targeted Drug Delivery

Pawan Kumar Pandey[1]    and Malay Kumar Das[2]

Department of Mechanical Engineering
Indian Institute of Technology Kanpur
Kanpur-208016, India

## Graphical Abstract

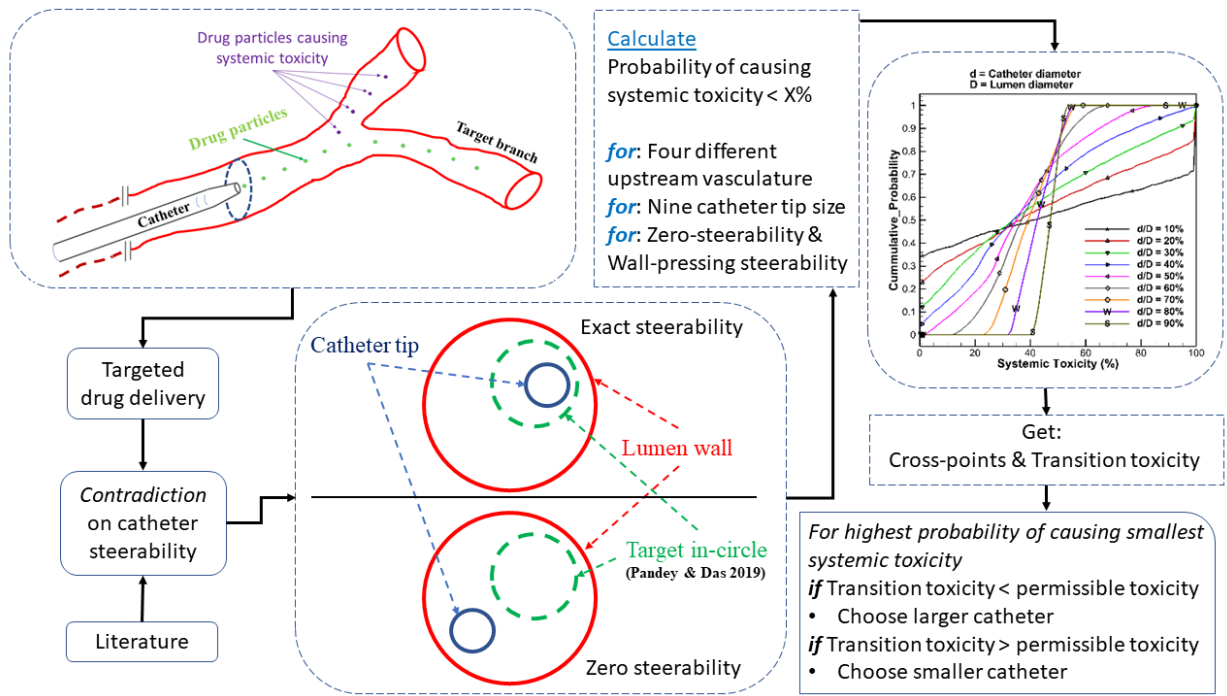

[1] Research Scholar, email: pawansciences@gmail.com, https://orcid.org/0000-0003-0145-3338
[2] Professor, email: mkdas@iitk.ac.in, https://orcid.org/0000-0002-7971-4166

## Abstract

In this work, we virtually study the intra-arterial targeted drug delivery. Specifically, this work models and quantifies the uncertainties associated with catheter steerability limitations. We classify catheter's limited steerability into two types, i.e., zero steerability, and wall pressing steerability. Further, we investigate the effects of steerability limitations on uncertainty of causing systemic toxicity levels, i.e., percentage of drug particles missing target. Proposed method quantifies the uncertainty of causing systemic toxicity in terms of probability. With this calculation approach, we look at the effects of upstream vasculature and catheter tip size. Results indicate the existence of a 'transition toxicity' level. Beyond transition toxicity level, larger catheters should be preferred over smaller catheters. Furthermore, we found that it is relatively easier to decide preferrable catheter size in zero-steerability than wall-pressing steerability conditions.

## 1. Introduction

Ideally, drug particles should only act on the diseased site. Practical challenges in drug delivery only allow for minimization of '*systemic toxicity*', i.e., drug particles acting on healthy tissues. Systemic toxicity of many drugs has fatal consequences. During delivery of such drugs, medical practitioner aim to achieve localized concentration of medication at the diseased site. '*Targeted drug delivery*' refers to the precise delivery of drug particles to the target site. Targeting mechanisms such as intrinsic physiological properties of circulation system, affinity of chemically engineered nanoparticles, and external guidance field such as magnetic field or ultrasound enable the precise delivery. Still, large percentage of drug particles miss the target site if not released in

Quantifying the Consequences of Catheter Steerability Limitations on Targeted Drug Delivery

proximity. Use of catheters, needles, microneedles, patch [1], inhalers, and micro-fluidic devices [2] minimize the distance between target and release site.

In present work, we study the intra-arterial drug delivery using catheters. Drug delivery through blood vessels find use in treatments of tumour, aneurysm, atherosclerosis, and clot dissolution [3–7]. These examples illustrate the potential of targeted drug delivery as medical therapy for several cardiovascular diseases. Hence, understanding of various aspects of intra-arterial targeted drug delivery has vital implications for successful treatment.

Treatment success require deposition of a sufficient amount of particles at the targeted site [8]. In parallel, prevention of side effects demand minimization of stray particles. Flow complications can limit the precision of drug delivery. Computer simulations address this limitation by calculating particle release map (PRM) [9]. PRM facilitates calculation of optimal catheter size, placement location and drug release strategy for minimization of systemic toxicity [10]. The strategy based on PRM assumes precise steerability of catheter tip. However, instrumentation limitations such as lack of catheter tip's steerability to target location within insertion plane can render PRM based planning ineffective. In this work, we look at the issue of catheter's steerability limitations.

In literature, many studies have used patient-specific and idealised geometries for targeted drug delivery investigations [11]. Challenges with targeted particle delivery arise from vasculature complications and particle characteristics [12]. Vascular complications such as tortuosity, three-dimensional twists, turns, and variations in the lumen area manifest prominently in cerebral arteries. As vascular complication increase, PRM (of different time instants) undergo significant changes. The frequent changes in PRM with time render ineffective the practice of catheter placement by first releasing dye.

Besides pulsatile flow, frequent changes in PRM find its root in vasculature induced flow unsteadiness. Bends in vasculature generate centrifugal forces in the flow, distorting the symmetric Womersley profile towards the bend's outer side [13,14]. Tortuous arteries have significant secondary flow [15], which influences flow structures and preferential branch selection [16]. Heavy particles with high velocity often exhibit tortuous trajectory. The highest momentum particle's gravitates toward the vessel's convex side [17]. Apart from density, particles' size and shape also play important role in determining their trajectory [18]. And since PRM are determined by backtracking the particle's trajectory, particle characteristics also influence the PRM.

Basciano et al. [9] demonstrated the PRM calculation for targeted delivery of microsphere particles in hepatic artery daughter vessels. However, their work uses idealistic geometry, and does not account for the twists present in the realistic vasculature. Childress and Kleinstreuer [19] reported a pseudo-dynamic approach for dynamic PRM determination. Though not tested for significant vasculature complexities, several investigators [20,21] have used a two-way modeling approach for the particle flow simulations. However, some results in the literature [21] show that for small to medium particle size, relative to lumen diameter, one-way coupling gives adequately good results as two-way coupling.

In literature, many works have investigated the properties of catheter and its impact on drug delivery. Sarkar et al. [22] presented the experience of choosing the best suiter catheter for angiography for percutaneous intervention in an anomalous artery. Classification based on the anatomical origin of the anomalous artery and iterative trial of the catheter was the basis of selecting an optimal catheter. Piper et al. [23] did a computational investigation on the effect of catheter size, tip position, infusion rate, and insertion angle on wall shear stress (WSS), particle residence time, blood cell damage, and stasis volume. Similarly, Sarker et al. [6] performed an optimization study for local drug infusion from the catheter to keep surrounding WSS under acceptable limits. Clark et al. [24] and Ararsa and Aldredge [25] investigated the effect of different types of catheter tip geometry on local hemodynamics, drug mixing, and thrombogenic potential responsible for catheter tip blockage. Ali et al. in [26] have documented several issues with catheter steerability along with its medical implications. Works of Richards et al. [8] and Anton et al. [27] have validated in-silico procedures with in-vitro and in-vivo results, respectively.

Most of the in-silico studies, on targeted drug delivery, make an implicit assumption that catheter tip can be accurately placed at the desired site. However, catheter steerability has limitations, and most of the studies do not take steerability limitations into account. Similarly, most of the studies have investigated the effect of particle characteristics and downstream vasculature on preferential branch selection. To the extent of our literature survey, we did not find any study considering the limitations of catheter steerability and upstream vasculature complexities. In this work, we are concerned about the uncertainties associated with the catheter tip placement [28]. A deterministic approach of finding catheter size and insertion position [10] assumes instrumental ability and precision to precisely position the catheter tip within the injection circle. Such precise control is hard to achieve and thus motivates this study to re-visit the problem of determining catheter size with a probabilistic method.

## 2. Methodology

Blood exhibits different behaviour in the different parts of vasculature. This complex behaviour arises from rheological properties and particulate nature of suspended cells. Both rheological and particulate behaviour have varying significance under different scales and pathological conditions [29]. In this work, studied cases mimic the scenario of drug delivery in cerebral artery. The relevant modelling and computational details are discussed in following subsections.

**2.1 Geometry:** Most in-silico investigations available in literature only take into account the downstream vascular complexities [8,9,19,30,31]. The upstream vasculatures can have significant influence on flow topology at the drug release plane. Flow topology and targeted drug delivery are inter-related [32]. With this knowledge as motivation, we select four arterial geometries. These geometries have identical downstream vasculature, i.e., a symmetric bifurcation. However, all four geometries have different upstream vasculatures. The upstream vasculatures vary in terms of twist. A twist of angles $180^0$, $360^0$, $720^0$, and $0^0$ is present in the approach part of geometry 1, 2, 3, and 4, respectively. Figure 1 show all four geometries along with the corresponding drug release plane.

**2.2 Assumptions and Problem Simplifications:** For the calculation, we assume free slip at the outer wall of the catheter. Drug injection rate is assumed to be the same as blood flow rate. Next, the effect of thickness of catheter wall is neglected in the present work. Arterial walls are assumed to be rigid.

**2.3 Governing Equations and Boundary Conditions:** In present work, blood flow is modeled as single-phase, incompressible, and Newtonian, see equation (1) and (2). We performed the necessary simulations using an finite volume method based in-house solver. To solve the mass and momentum conservation equations, solver uses a modified SIMPLE algorithm [33].

$$\frac{\partial u_i}{\partial x_i} = 0 \qquad (1)$$

$$\frac{\partial u_i}{\partial t} + u_j \frac{\partial u_i}{\partial x_j} = -\frac{1}{\rho} \frac{\partial p}{\partial x_i} + \frac{1}{\rho} \frac{\partial}{\partial x_j} \left[ \mu \left( \frac{\partial u_i}{\partial x_j} + \frac{\partial u_j}{\partial x_i} \right) \right] \qquad (2)$$

The equation (1) and (2) are numerically solved along with reality mimicking boundary conditions. At inlet, Womersley profile, i.e., fully developed velocity profile under pulsatile flow is imposed. The Womersley profile is derived from average velocity waveform shown in Fig. 2(a). Next, no-slip boundary condition is imposed at the arterial wall. We performed computations until consecutive three waveform cycles are free from any phase lag.

**2.4 Rheological Model:** Blood starts to exhibit shear-thinning behavior as it flows major arteries to smaller arteries. The generalized non-Newtonian models can model the shear thinning behavior [34]. Non-Newtonian behavior gain significance in slow flow regions [35]. In present work, besides investigating the effect of catheter steerability we also look at the effect of upstream vasculature induced helical flow. The helical flow or secondary flow can get influenced by the shear-thinning model [36]. Therefore, in present work, we intentionally take viscosity as constant, i.e., infinite shear viscosity (0.0035 Pa-s). The Newtonian assumption allows us to distill the effect of upstream twist and tortuosity on particle release maps.

**2.5 Particle Transport:** Particle transport is carried out using a one-way coupling approach. A small-time step of size $10^{-4}$ sec is used to get a smooth trajectory. Equation 3 represents the dynamics of the particle motion. Subscript 'p' refers to particle properties, while the right side has a summation of all force terms.

$$m_P \frac{dv_P}{dt} = \sum_i F_{P,i}$$ (3)

**2.6 Code Validations:** The in-house solver, used for simulations in present work, well tested, and validated [10,37–41]. Additionally, in Fig. 2(b) &(c), validation of pulsatile flow in a curved tube is validated against the computational and experimental results of Timite et al. [42].

**2.7 Computational Pipeline & Data Analysis:** At first, we did flow simulations in all four geometries. Once we finish the calculation of primitive variables, the dynamic particle release map (DPRM) is calculated at the injection plane. Details of the DPRM calculation procedure can be found in our previous work [10]. The computational pipeline's end goal is to calculate the probability of causing a certain level of systemic toxicity for a chosen catheter size. A schematic of multiple steps in the computational pipeline is shown in Fig. 3(a).

After DPRM calculation, numerous (~4000 in present work) *sample circles* are placed over the injection plane. While sample circles crossing the periphery of the injection place are declared invalid, others are designated as *valid sample circles*. These sample circles can be understood as the probable position of the catheter mouth. For each valid sample circles, systemic toxicity is calculated as follows:

$$\text{Systemic Toxicity} = \frac{N - Nt}{N} \qquad (4)$$

where N = Total number of DPRM points within the sample circle

and Nt = Number of DPRM points for target branch

Systemic toxicity value for each valid sample circle is mapped to circle's center, thus giving systemic toxicity contours. Edges of systemic toxicity contour and injection plane differ by the radius of the sample circle radius. Now after calculating the probability of causing systemic toxicity, in a specified range, for a given catheter size (same as sample circle) is calculated using the following formula:

$$P = \frac{A1}{A} \qquad (5)$$

where P = probability of causing systemic toxicity less than Y% and more than X%

A1 = Contour area with systemic toxicity less than Y% and more than X%

A = Total area of systemic toxicity contour

**2.8 Types of Catheter Steerability:** Depending upon the technology available to the medical practitioner, the deviation of catheter placement from the intended position can be significant. We classify catheter steerability in the following four types:

a) *Exact Steerability*: Catheter center can be matched with the desired points within the lumen. Catheter deviations << lumen radius.

b) *Wall Pressing Steerability*: Catheter wall would be touching the lumen wall. However, azimuthal position control is absent.

c) *Partial Steerability*: Catheter center can not be matched exactly with the desired points within the lumen. However, deviations are less than the lumen diameter. Catheter deviations < lumen radius.

d) *Zero Steerability*: There is absolutely no control on the catheter position in the lumen cross-section. Catheter deviation = lumen radius.

Our previous work [10] has discussed the different drug release strategies assuming *Exact Steerability*. In the present work, we present and discuss the other catheter steerability. Wall pressing steerability and zero steerability is illustrated in Figure 3(b).

## 3. Results & Discussions

In the present work, we study the DPRMs and perform further analysis based on DPRMs calculated for tracer particles. Flow in geometry-4 does not provide any conditions enabling preferential branch selection. Therefore, the DPRM of geometry 4 is symmetric to the plane of bifurcation, see Fig. 4(a). Figure 4(a) shows the dynamic particle release map (DPRM) for geometry-4. Because of symmetric and identical DPRM throughout pulse's time period, we show its single instance, i.e., at t=0.01 sec. Additionally, Figure 4(b) shows the corresponding cumulative probability distribution plot for zero steerability of nine catheters of different sizes. All of the cumulative probability curves are found to intersect at single point. These intersection points of 'S' shaped cumulative probability lines are termed as *cross points*. There is only one cross point in the case of symmetric upstream vasculature and tracer particles. Toxicity value corresponding to a cross-point is defined as transition toxicity. The transition toxicity, in geometry 4, is at 50%. Figure 4(b) visually illustrates the cross points and transition toxicity values.

On the contrary with geometry-4, in geometry-1, 2 & 3, DPRMs are not symmetric; see Fig. 5. In this work, we calculate and present DPRM for nine equally distant time instants over the waveform pulse, see Fig. 2(a). In all three geometries 1, 2 & 3, with tortuous upstream vasculature, DPRM zones are splintered, and zone boundaries are irregular. In geometry 1, primarily, both DPRM-zones are simply-connected, except a few islands of zone-1, and the effect of upstream tortuosity reflects only in the distorted interface of both DPRM zones. In both geometries 2 & 3, none of the DPRM-zones are simply-connected. In geometry-2 results, DPRM-zone1 is larger than DPRM-zone2; the reverse is valid for geometry-3. Significant changes in the shape of DPRM zones happen only during the systole phase of the waveform. DPRM zone shapes undergo more changes, over the waveform time period, in geometry-1 than geometries-2 & 3. From a purely geometrical point of view, DPRMs of geometries 2 &3 are more similar than with DPRMs of geometry 1. After doing suitable rotational transformation, the similarity of DPRMs of geometries 2 & 3 can be shown. However, the preferred branch is different in both geometries.

Figure 6 shows the probabilities of causing a different range of systemic toxicity by nine catheters with zero steerability. With a bin size of 10%, the bar plot is used to show the probability distribution at around the systole peak (t=0.11 sec). The nine catheter sizes are 10%, 20%, 30%, 40%, 50%, 60%, 70%, 80% & 90% of lumen diameter. Small catheters, with diameter up to 30% of lumen diameter, show a higher probability of causing either very low, i.e., 0-10%, or very high, i.e., 90-100% systemic toxicity. Catheters with a diameter larger than 40% of lumen diameter show a higher probability of causing an intermediate level of systemic toxicity and are less likely to cause very low or very high systemic toxicity. Behavior-wise, the smallest three catheters (10%, 20% & 30% of lumen diameter) exhibit u-shaped probability distribution. Large catheter diameters (d/D=50% to 90%), on the other hand, exhibit a plateau-shaped probability distribution. . For geometry-2 (Fig. 6(b)) and geometry-3 (Fig. 6(c)), Two catheter sizes, i.e., d/D=10% & 20% show u-shape distribution and six larger catheter sizes, i.e., d/D=40%, 50%, 60%, 70%, 80% and 90% show bell-shaped distribution. Catheter size d/D=30% show uniform distribution.

Figure 7 shows the cumulative probability distribution plots for all nine catheters, with zero steerability, in geometries 1, 2 & 3. For brevity, these results are shown only for two different time instants i.e., at t=0.01, 0.11 sec. All catheters with a diameter of less than 50% of lumen diameter have a nonzero probability of causing zero toxicity. This plot enables us to see that while larger catheter sizes limit the range of minimum and maximum possible toxicity, smaller catheter sizes allow the possibility of zero toxicity. The plots show that cross-points are scattered in between 25-55%, 25-45%, and 45-75% of systemic toxicity in geometry-1, 2, and 3, respectively. These scattered cross points or corresponding range of transition toxicity can be of significant help while choosing the catheter for the medical interventions. If permissible toxicity is larger than transition toxicity, then opting large catheter size would be beneficial because a large catheter would have a higher probability of causing systemic toxicity within the acceptable limit.  However, if permissible toxicity is smaller than transition toxicity, choosing a smaller catheter size has more chances of keeping toxicity levels within permissible levels.

Figure 8 presents the cumulative probability distribution plot for wall pressing steerability in geometry 1, 2, and 3. A quick comparison between Fig. 7 and 8 shows that cumulative probability plots of wall-press steerability are not as smooth as of zero steerability. Moreover, the curve does not necessarily exhibit 'S' shaped behavior, and more than one inflection point exists.

*Steady phase equivalent calculations*

Performing multi-modal pulsatile flow simulations is computationally intensive and requires a large amount of time to do calculations. However, the end-users may not have such advanced computational capacities to perform such calculations in a reasonable time period. So, we evaluate an alternate approach of calculation, i.e., steady phase equivalent approach. In this method, instead of performing the complete pulsatile flow simulation, we perform several steady-state simulations. At the inlet, inflow velocity is given according to the pulsatile flow waveform of the corresponding time instant. Similar to unsteady cases, particle pathline calculations and subsequent analysis are also done for the steady phase equivalent approach.

Calculated steady PRMs, alongside DPRM, for geometries 1, 2 & 3 are shown in Figure 9. For geometry-1, there are significant differences in the PRM shape of the unsteady and steady phase equivalent cases. For geometry-2 and geometry-3, DPRMs and steady-PRM are similar to each other. Figure 10 shows that, in geometry 1 & 2, the steady phase equivalent approach under-predicts the probability of causing lower systemic toxicity levels and over-predicts the probability of causing higher systemic toxicity levels; vice-versa is observed for geometry 3. Errors in probability prediction depend on the catheter size; as size increases, error increases. One factor to consider here is that the larger the catheter size gets smaller its range of probable systemic toxicity. So for most systemic toxicity intervals towards higher or lower extremities, larger catheter sizes have zero probability. Therefore, errors are large and contained only in the plateau region of the probability distribution.

Figure 11 (a), (c), (e) show all cross points corresponding to nine catheters, at various time instants, in geometry 1, 2, and 3, respectively. Figure 11 (b), (d), (f) show cross points with minimum and maximum transition toxicity levels. Results show, for geometry-1, cross points are scattered between 22% to 55% systemic toxicity. With an increase in the twist in geometry-2, having $360^0$ twists, transition toxicity levels slide on the lower side and are scattered between 22% to 50% systemic toxicity. In contrast to geometry-2, a further increase in the arterial twist in geometry-3 leads to shifting transition toxicity levels towards higher sides and scattered between 47% to 77% systemic toxicity. If the permissible level of systemic toxicity is less than 22% in geometry-1, then choosing a small catheter size creates more possibility of causing systemic toxicity within the permissible limit. If the permissible level of systemic toxicity is more than 55%, then choosing a larger catheter size would lead to a higher chance of containing systemic toxicity within the permissible limit. However, if the permissible level of systemic toxicity is between 22% to 55%, then the choice of catheter should be made only after detailed analysis based on all cross-points. Similar logic holds for geometry 2 and 3 for lower transition toxicity levels of 22% and 47% and higher transition toxicity levels of 50% and 77%, respectively.

Figure 12 shows the comparison of cross point predictions from unsteady calculations and steady phase equivalent calculations. For geometry-1, Cross-point predictions from the steady phase equivalent approach differ significantly from the prediction of unsteady simulations. However, for geometry-2 and geometry-3, differences are not significant. Therefore, within the constraints of limited results available to us, we conclude that the steady phase equivalent approach could provide sufficient accurate predictions for geometries with a significant upstream twist as in geometry-2 ($360^0$) and geometry-3 ($720^0$).

The cumulative probability plot for wall pressing steerability is shown in Figure 13. Since wall pressing steerability has an inherently higher degree of control over catheter position than in zero steerability. Therefore, intuitively we were expecting cross points of wall steerability to be less scattered than zero steerability results. However, as seen in Fig. 8 and Fig. 13, obtained distribution of cross-points is counter-intuitive. More dispersed PRM on the outer ring than inner rings of the injection plane is the reason for this behavior. However, upon comparing the results of both steerability, we find that it is more difficult to choose optimum catheter size for wall pressing steerability than zero-steerability conditions.

## 4.  Conclusions

This work modeled and studied the targeted drug delivery using catheters with limited or zero steerability. Investigated geometries have identical downstream bifurcation but different upstream vasculature. The upstream twist generates a swirl at the injection plane. The difference in flow at drug injection plane results in different DPRM. Catheter steerability limitations cause uncertainty in systemic toxicity. Present work calculated the corresponding probabilities of causing systemic toxicity. For analysis, this work introduced the concepts of cross points and transition toxicity. Some specific conclusions according to relevance are listed as following:

*Of computational relevance*

i.    With an increase in the twist of upstream vasculature, DPRM zones become non-simply-connected and splintered.

ii.   The performance of full unsteady approach and steady phase equivalent approach is investigated in terms of introduced concepts of cross points and transition toxicity. The steady phase equivalent approach gives better results for geometries with higher upstream tortuosity. This is due to the ability of helicity, induced due to upstream twist, to resist the disturbance in flow structures due to physiological pulsation.

*Of biomedical relevance*

iii.  Results indicate the existence of 'transition toxicity' above which larger catheter size would serve better than smaller catheter sizes. Therefore, knowledge of 'transition toxicity' beforehand can be immensely helpful for medical practitioners while choosing catheter size.

iv. Wall pressing steerability results are counter intuitive. The corresponding scattered cross points complicate the choice between smaller and larger catheter diameters for ensuring the permissible level of systemic toxicity.

**Acknowledgment**

We gratefully acknowledge the computing facilities provided by the Indian Institute of Technology Kanpur.

Quantifying the Consequences of Catheter Steerability Limitations on Targeted Drug Delivery

Release 2016;238:149–56. doi:10.1016/j.jconrel.2016.07.041.

[8]     Richards AL, Kleinstreuer C, Kennedy AS, Childress E, Buckner GD. Experimental microsphere targeting in a representative hepatic artery system. IEEE Trans Biomed Eng 2012;59:198–204. doi:10.1109/TBME.2011.2170195.

[9]     Basciano CA, Kleinstreuer C, Kennedy AS, Dezarn WA, Childress E. Computer modeling of controlled microsphere release and targeting in a representative hepatic artery system. Ann Biomed Eng 2010;38:1862–79. doi:10.1007/s10439-010-9955-z.

[10]    Pandey PK, Das MK. Unsteady Targeted Particle Delivery in Three Dimensional Tortuous Cerebral Artery. Int J Adv Eng Sci Appl Math 2019;11:263–79. doi:10.1007/s12572-020-00263-9.

[11]    Malone F, McCarthy E, Delassus P, Buhk JH, Fiehler J, Morris L. Investigation of the Hemodynamics Influencing Emboli Trajectories Through a Patient-Specific Aortic Arch Model. Stroke 2019;50:1531–8. doi:10.1161/STROKEAHA.118.023581.

[12]    Waterman KC, Sutton SC. A computational model for particle size influence on drug absorption during controlled-release colonic delivery. J Control Release 2003;86:293–304. doi:10.1016/S0168-3659(02)00418-2.

[13]    Gijsen FJ, van de Vosse FN, Janssen JD. The influence of the non-Newtonian properties of blood on the flow in large arteries: steady flow in a carotid bifurcation model. J Biomech 1999;32:601–8. doi:10.1016/S0021-9290(99)00015-9.

[14]    Timité B, Castelain C, Peerhossaini H. Pulsatile viscous flow in a curved pipe: Effects of pulsation on the development of secondary flow. Int J Heat Fluid Flow 2010.

**List of Figures:**

Figure 1.  Geometries used in the study (a) Geometry 1 has $180^0$ of upstream twist; (b) geometry 2 has $360^0$ of upstream twist; (c) geometry 3 has upstream twist of $720^0$ twist; (d) geometry 4 has straight tube as approach tube. Lumen cross-section shown with velocity contour is the plane of catheter placement. The downstream bifurcation in all four geometries is symmetric and identical.

Figure 2.  (a) Flow waveform applied at the inlet using Womersley profile. Symbols indicate time instants for which DPRM are calculated. (b)-(c) Velocity profile of three different phases of pulsatile flow in curved tube is compared with experimental and numerical results of Timite et al. 2010.

Figure 3.  (a) Calculation Procedure of Probability of causing systemic toxicity less than X%. The illustrated procedures correspond to zero steerability condition – meaning catheter tip could be anywhere in the injection plane. (b) Visual illustration of possible locations of catheter tip in different types of catheter steerability.

Figure 4.  (a) Dynamic particle release map for geometry 4. Blue symbols indicate particle release positions which lead to the left daughter branch. Green symbols indicate the particle release position which lead to the target, i.e., right daughter branch. Red symbols indicate the site from where released particles either stick to wall or fail to enter any branch during three waveform time periods from release. (b) Cumulative probability plot over complete range of systemic toxicity for nine different catheter sizes. Please note: the visual illustration of newly defined quantities, i.e., Transition toxicity and cross points.

Figure 5.  Dynamic particle release map (DPRM) for tracer particles at t=0.01, 0.11, 0.21, 0.31, 0.41, 0.51, 0.61, 0.71, 0.81 sec (row-wise) for geometry-1 (first column), geometry-

2 (second column) and geometry-3 (third column). Green colored points indicate location which lead to target branch, i.e., right branch.

Figure 6.   Histogram plot of probability against systemic toxicity over range of 0 to 100% with interval size of 10%. Results are shown for unsteady flow conditions in geometry-1, 2 and 3 using nine different catheter sizes. These results correspond to the case of zero steerability of catheters at t=0.11 seconds.

Figure 7.   Plots showing cumulative probability of causing systemic toxicity from 0 to 100% for nine different catheter sizes. Results shown are for case of zero steerability of catheters in the unsteady flow conditions. These results are shown for two-time instants, i.e., t=0.01 & 0.11 seconds.

Figure 8.   Plots showing cumulative probability of causing systemic toxicity from 0 to 100% for nine different catheter sizes. Results shown are for case of wall pressing steerability of catheters in the unsteady flow conditions. These results are shown for two-time instants, i.e., t=0.01 & 0.11 seconds.

Figure 9.   Comparison of unsteady and steady phase equivalent particle release maps (PRM) for geometry-1, 2 & 3. Dynamic particle release map (DRPM) and steady PRM (at t= 0.11 and t=0.41 seconds) of geometry-1,2, and 3 are compared.

Figure 10. Error in probability estimation using steady phase equivalent approach. Results shown are for systemic toxicity produced by nine different catheter sizes with zero steerability. (a) geometry-1 at t=0.01 sec (b) geometry-1 at t=0.41 sec (c) geometry-2 at t=0.01 sec (d) geometry-2 at t=0.41 sec (e) geometry-3 at t=0.01 sec (f) geometry-3 at t=0.41 sec

Figure 11.  Cross points for all three unsteady cases and zero steerability of catheters. All cross points are shown in (a) for geometry-1 (c) for geometry-2 (e) for geometry-3. Maxima and minima of cross points (to indicate the range) is shown in (b) for geometry-1 (d) for geometry-2 and (f) for geometry-3.

Figure 12.  Cross points comparisons for (a) all unsteady case results in geometry 1, 2 and 3; (b) for geometry-1 (c) for geometry-2 (d) for geometry-3 with unsteady and steady phase equivalent estimation. These results are of zero-steerability of catheters.

Figure 13.  Comparison of cross points in case of zero steerability and wall pressing steerability. Results are shown for (a) geometry-1 (b) geometry-2 and (c) geometry-3.

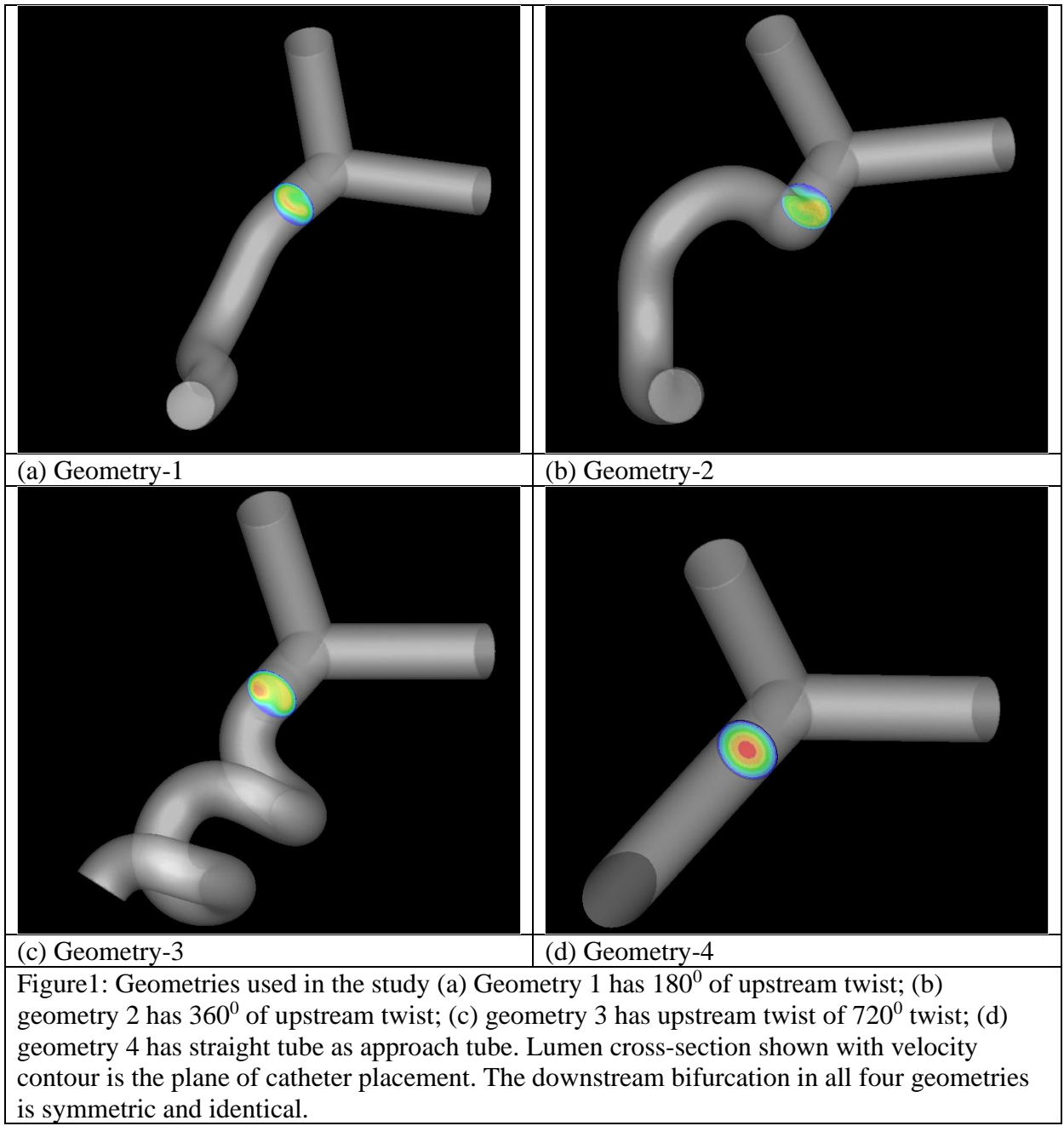

(a) Geometry-1

(b) Geometry-2

(c) Geometry-3

(d) Geometry-4

Figure1: Geometries used in the study (a) Geometry 1 has $180^0$ of upstream twist; (b) geometry 2 has $360^0$ of upstream twist; (c) geometry 3 has upstream twist of $720^0$ twist; (d) geometry 4 has straight tube as approach tube. Lumen cross-section shown with velocity contour is the plane of catheter placement. The downstream bifurcation in all four geometries is symmetric and identical.

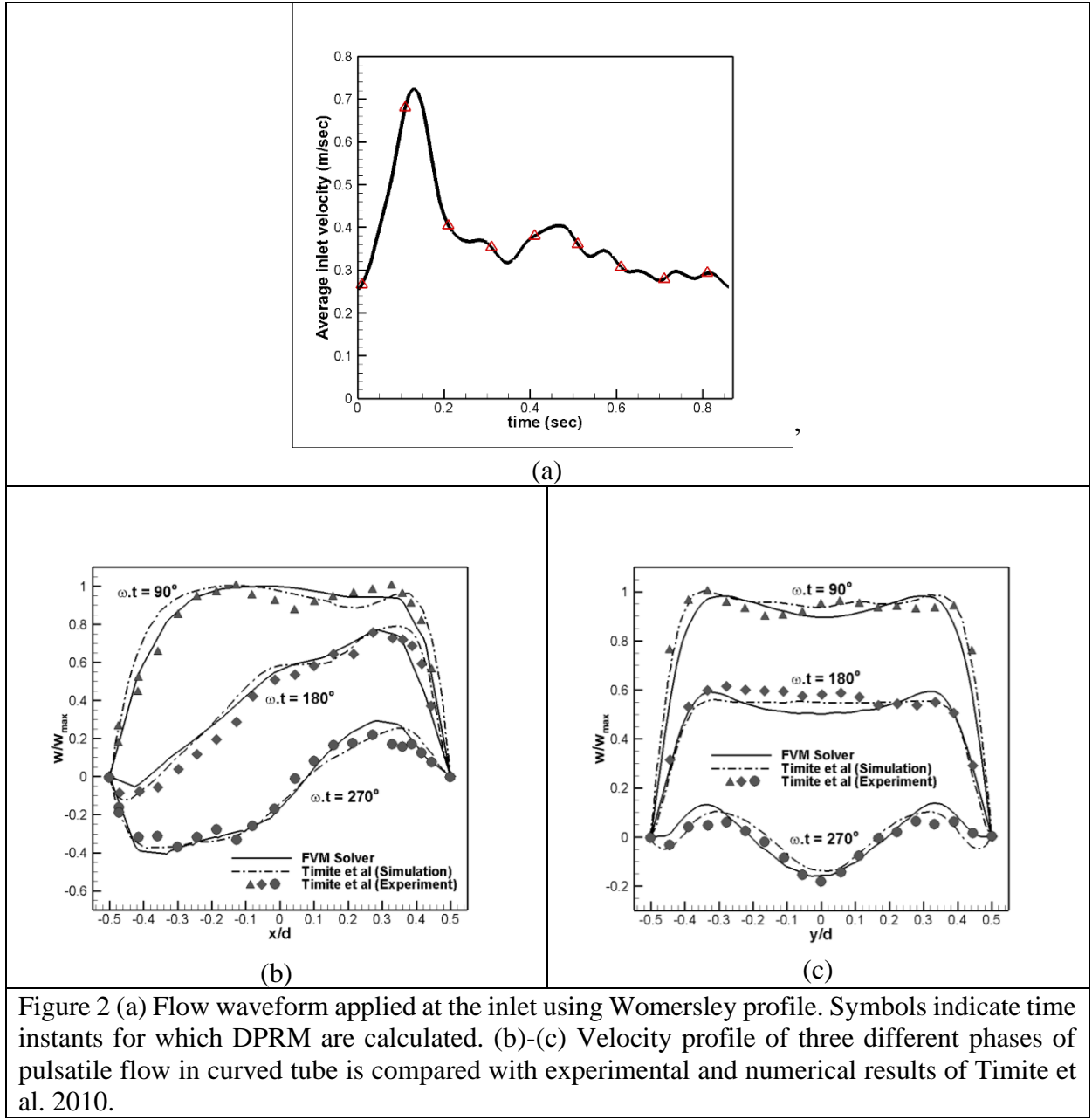

Figure 2 (a) Flow waveform applied at the inlet using Womersley profile. Symbols indicate time instants for which DPRM are calculated. (b)-(c) Velocity profile of three different phases of pulsatile flow in curved tube is compared with experimental and numerical results of Timite et al. 2010.

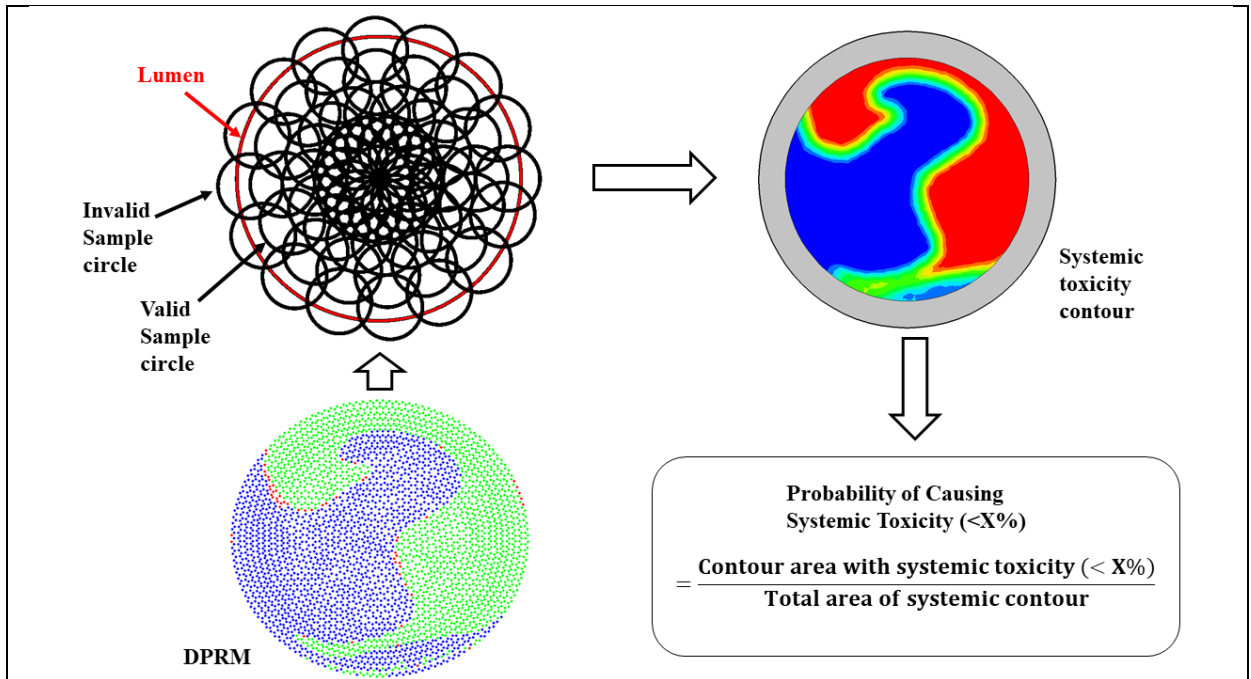

(a)

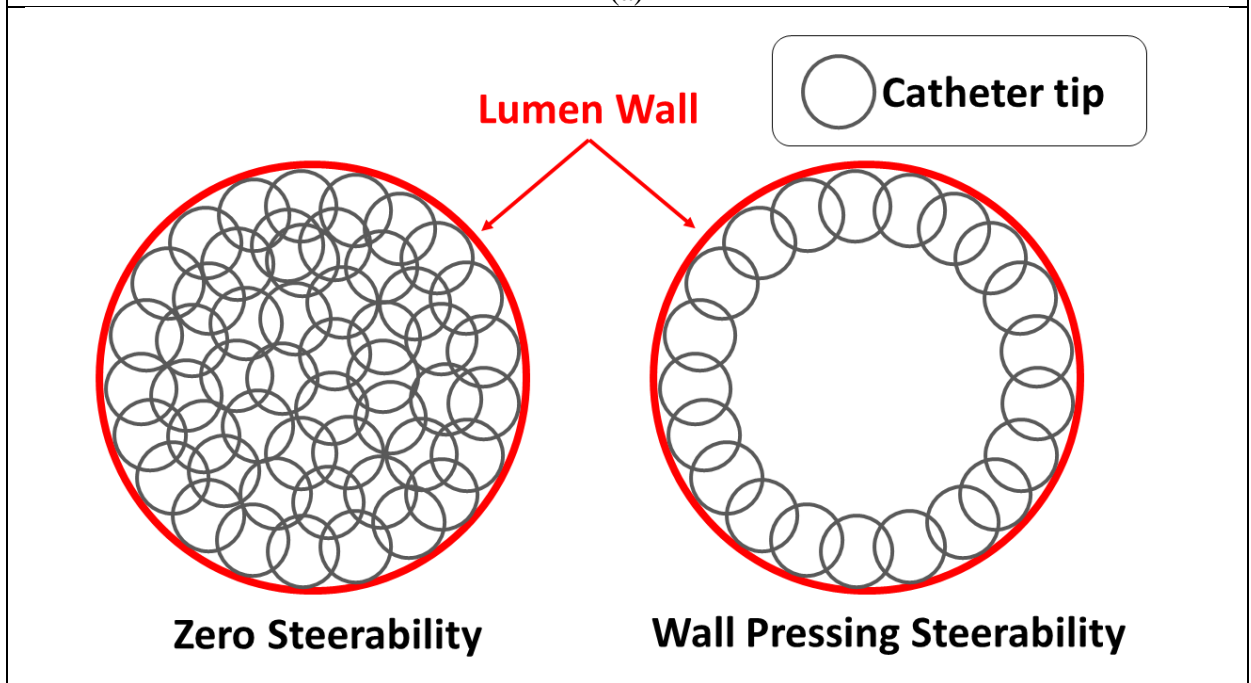

(b)

Figure 3: (a) Calculation Procedure of Probability of causing systemic toxicity less than X%. The illustrated procedures correspond to zero steerability condition – meaning catheter tip could be anywhere in the injection plane. (b) Visual illustration of possible locations of catheter tip in different types of catheter steerability.

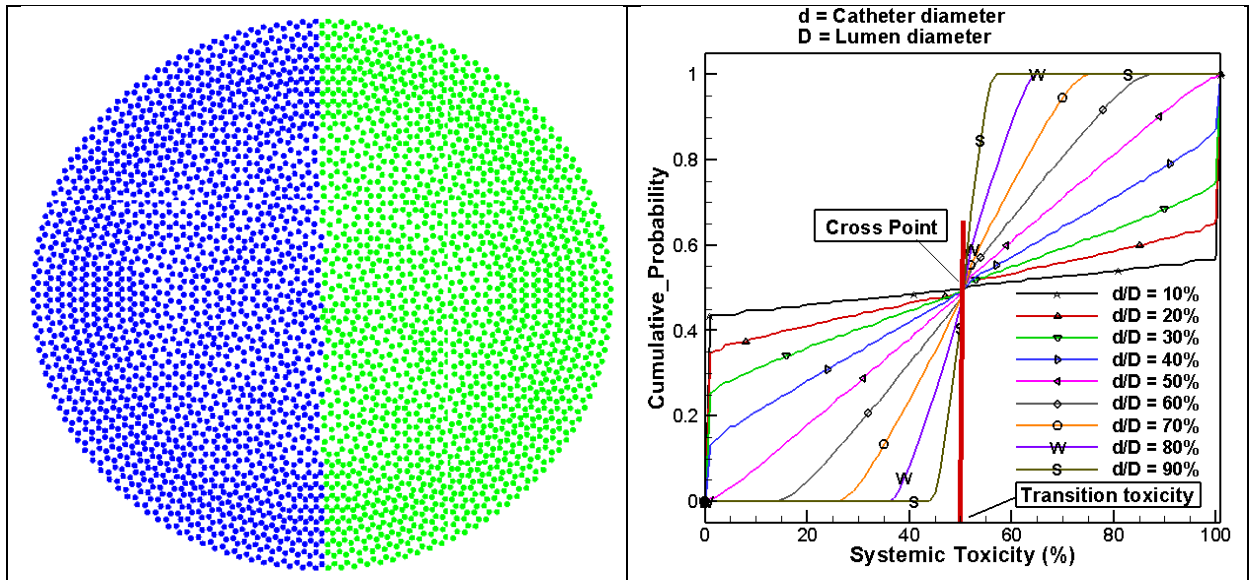

(a)

(b)

Figure 4: (a) Dynamic particle release map for geometry 4. Blue symbols indicate particle release positions which lead to the left daughter branch. **Green symbols indicate the particle release position which lead to the target, i.e., right daughter branch**. Red symbols indicate the site from where released particles either stick to wall or fail to enter any branch during three waveform time periods from release. (b) Cumulative probability plot over complete range of systemic toxicity for nine different catheter sizes. Please note: the visual illustration of newly defined quantities, i.e., Transition toxicity and cross points.

| Geometry 1 | Geometry 2 | Geometry 3 |
|---|---|---|

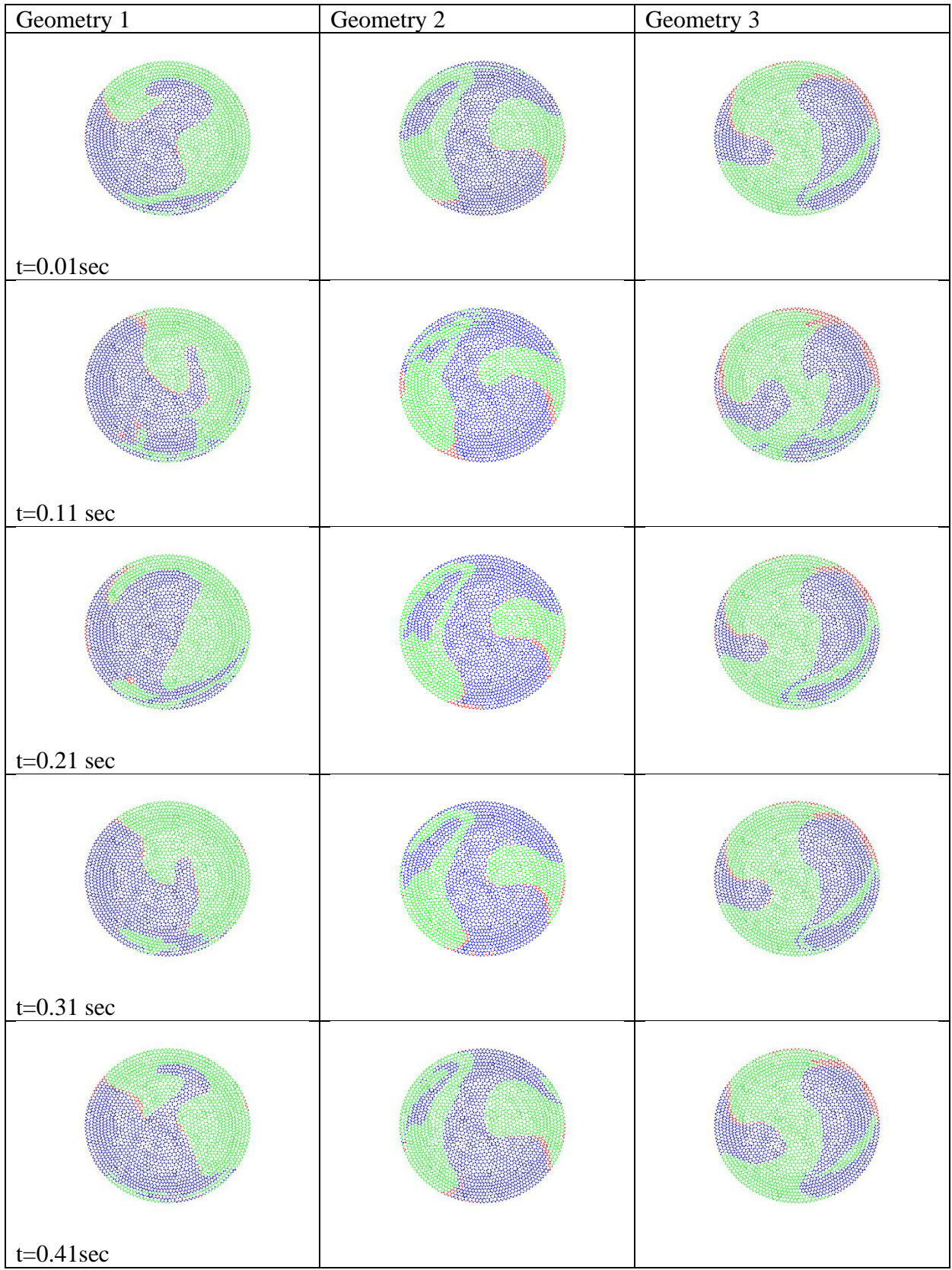

t=0.01sec

t=0.11 sec

t=0.21 sec

t=0.31 sec

t=0.41sec

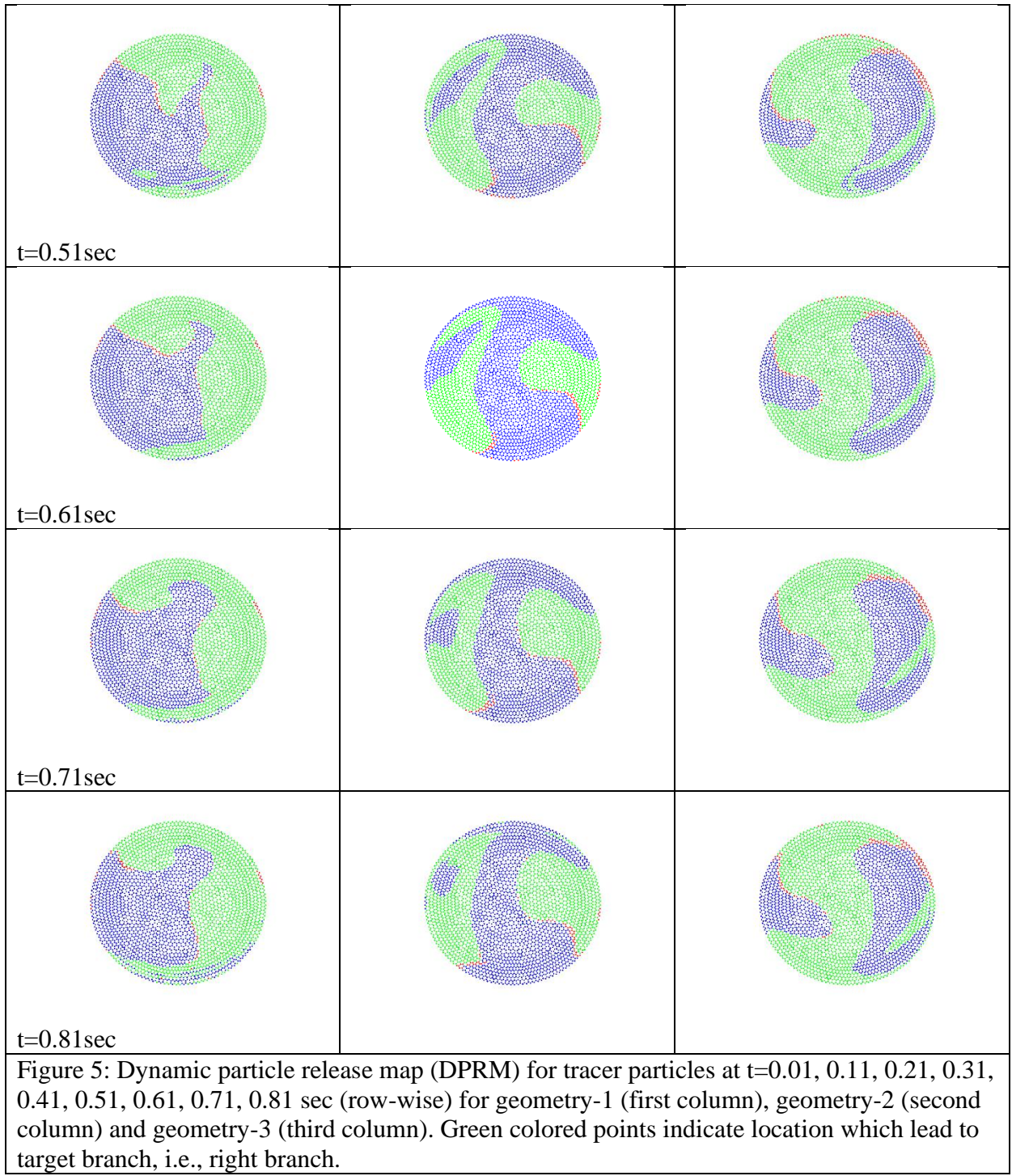

Figure 5: Dynamic particle release map (DPRM) for tracer particles at t=0.01, 0.11, 0.21, 0.31, 0.41, 0.51, 0.61, 0.71, 0.81 sec (row-wise) for geometry-1 (first column), geometry-2 (second column) and geometry-3 (third column). Green colored points indicate location which lead to target branch, i.e., right branch.

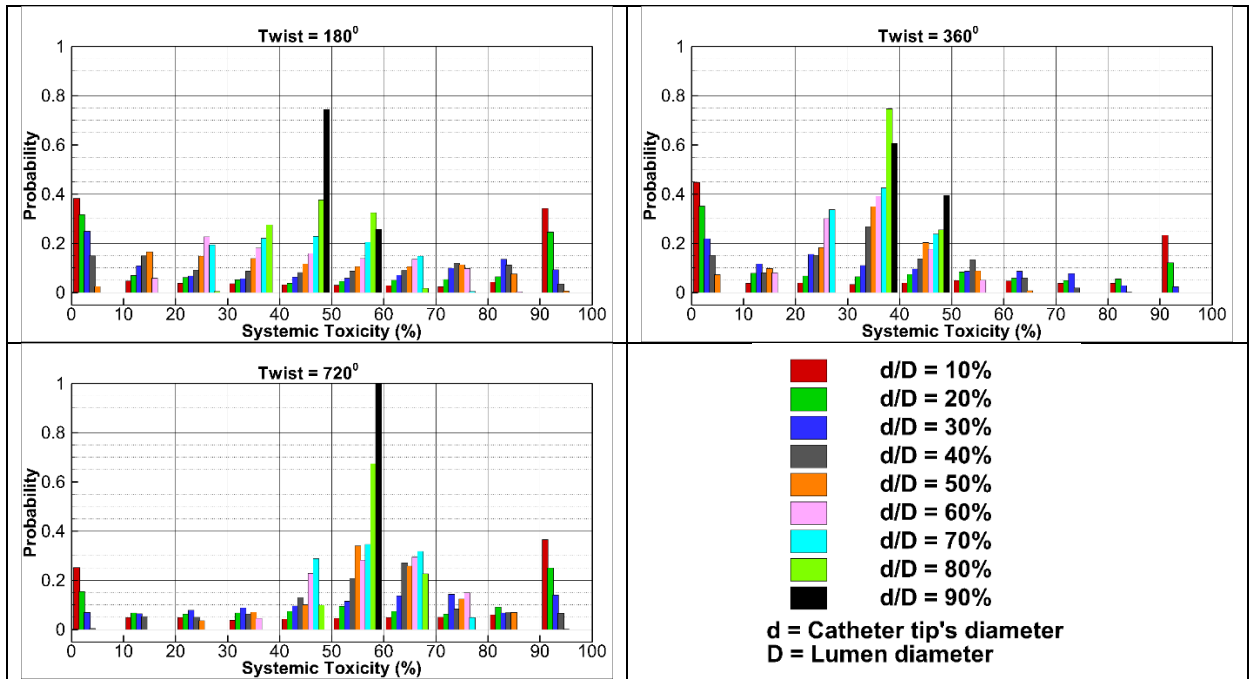

Figure 6: Histogram plot of probability against systemic toxicity over range of 0 to 100% with interval size of 10%. Results are shown for unsteady flow conditions in geometry-1, 2 and 3 using nine different catheter sizes. These results correspond to the case of zero steerability of catheters at t=0.11 seconds.

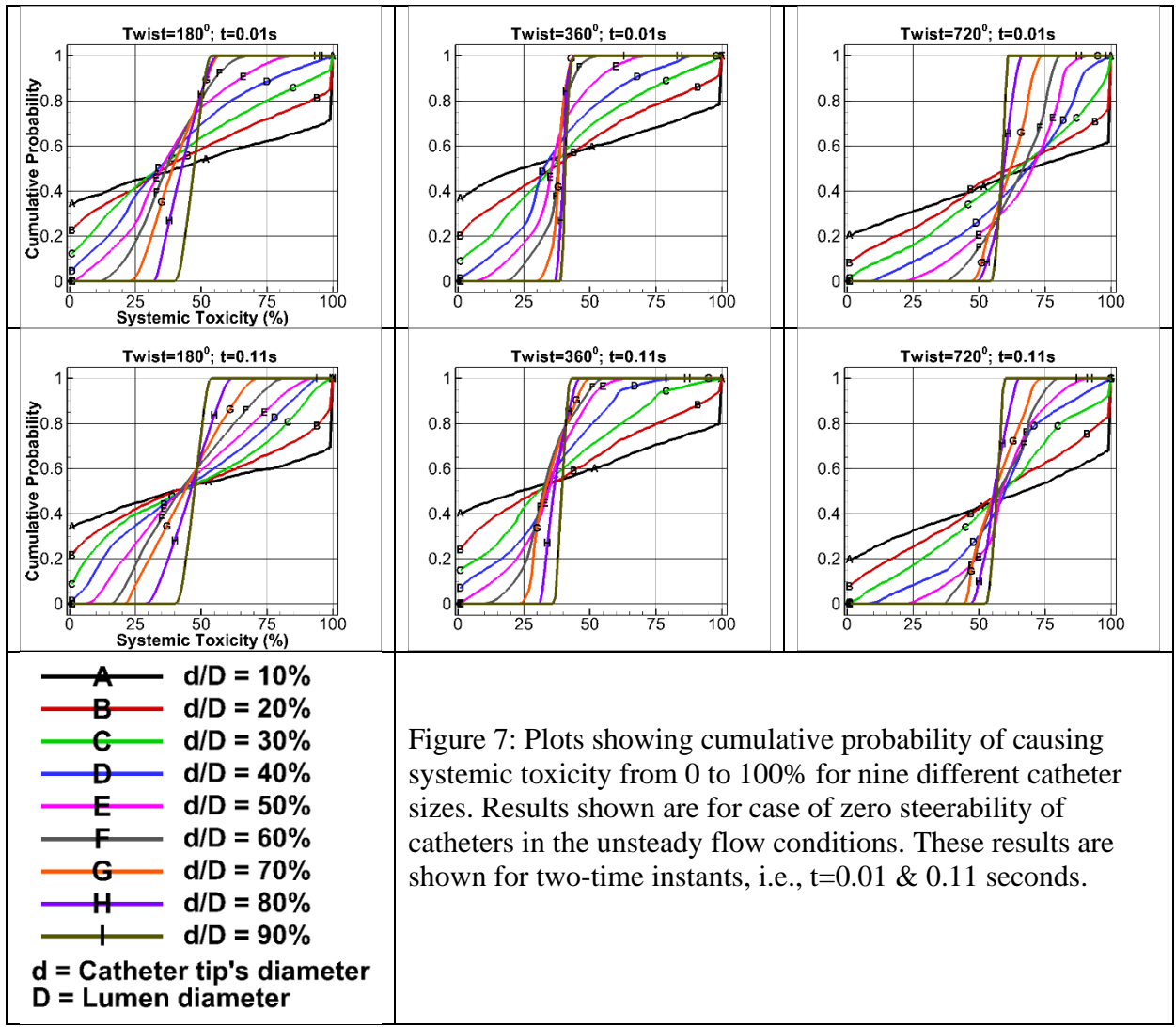

Figure 7: Plots showing cumulative probability of causing systemic toxicity from 0 to 100% for nine different catheter sizes. Results shown are for case of zero steerability of catheters in the unsteady flow conditions. These results are shown for two-time instants, i.e., t=0.01 & 0.11 seconds.

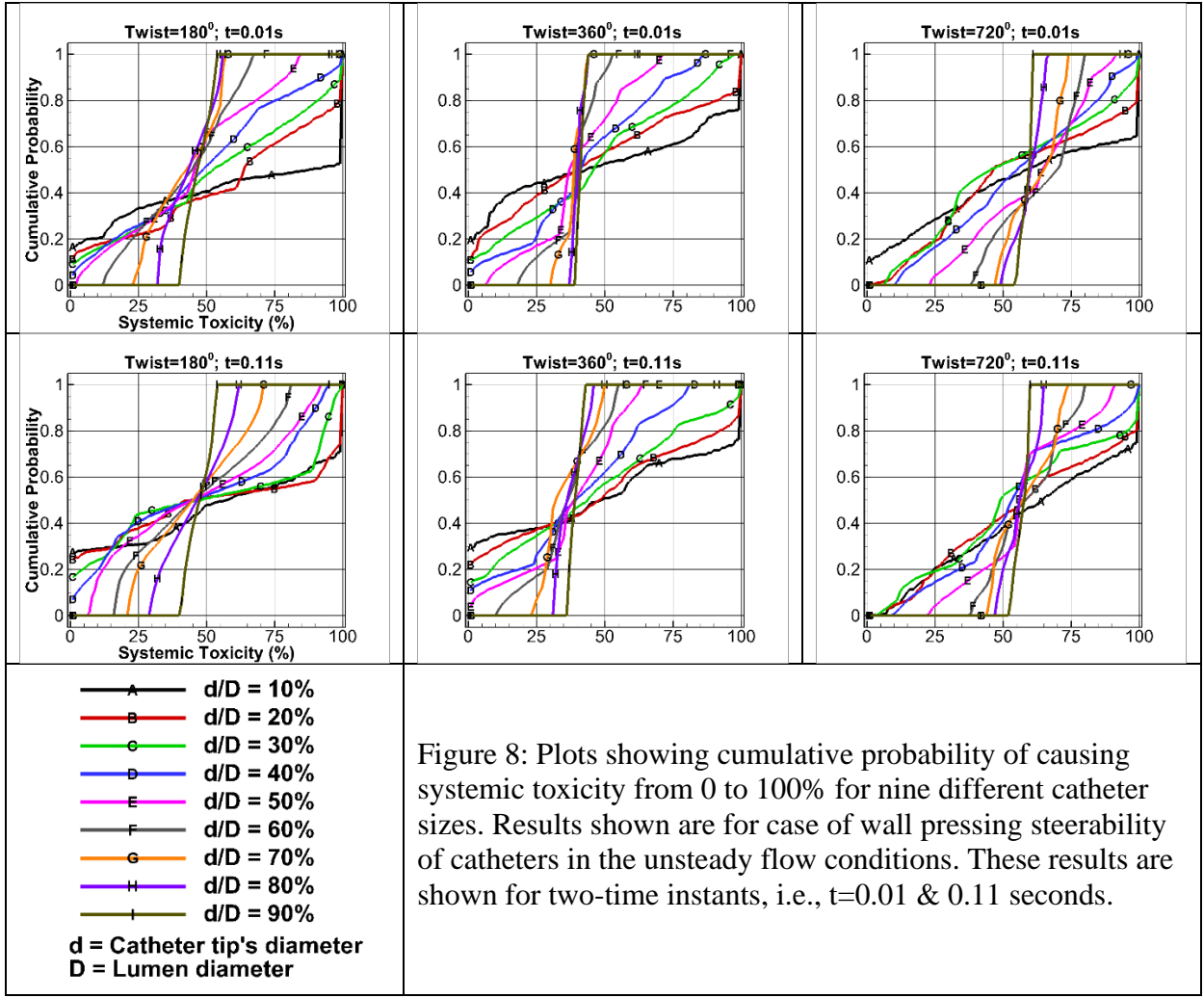

Figure 8: Plots showing cumulative probability of causing systemic toxicity from 0 to 100% for nine different catheter sizes. Results shown are for case of wall pressing steerability of catheters in the unsteady flow conditions. These results are shown for two-time instants, i.e., t=0.01 & 0.11 seconds.

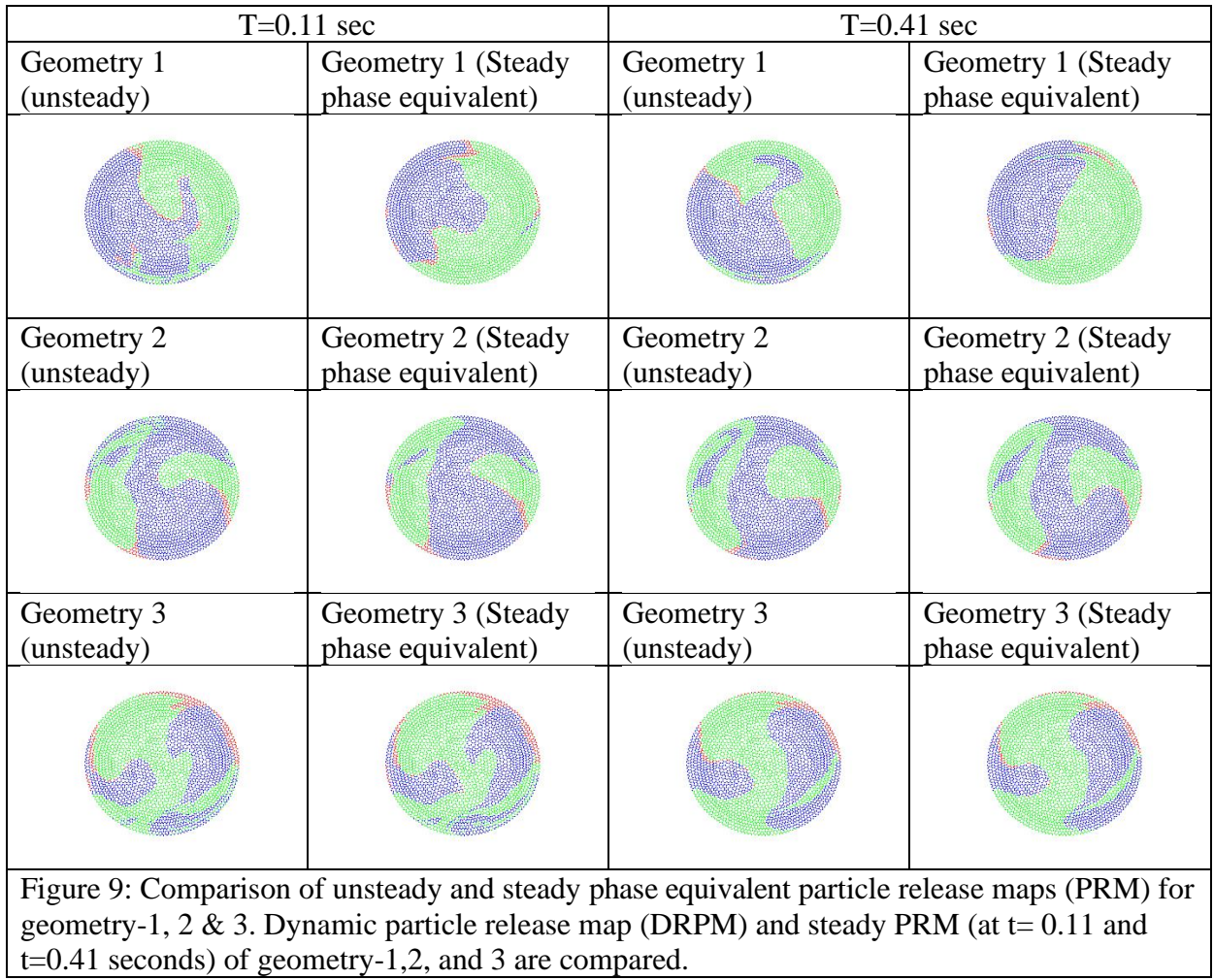

| T=0.11 sec | | T=0.41 sec | |
|---|---|---|---|
| Geometry 1 (unsteady) | Geometry 1 (Steady phase equivalent) | Geometry 1 (unsteady) | Geometry 1 (Steady phase equivalent) |
| Geometry 2 (unsteady) | Geometry 2 (Steady phase equivalent) | Geometry 2 (unsteady) | Geometry 2 (Steady phase equivalent) |
| Geometry 3 (unsteady) | Geometry 3 (Steady phase equivalent) | Geometry 3 (unsteady) | Geometry 3 (Steady phase equivalent) |

Figure 9: Comparison of unsteady and steady phase equivalent particle release maps (PRM) for geometry-1, 2 & 3. Dynamic particle release map (DRPM) and steady PRM (at t= 0.11 and t=0.41 seconds) of geometry-1,2, and 3 are compared.

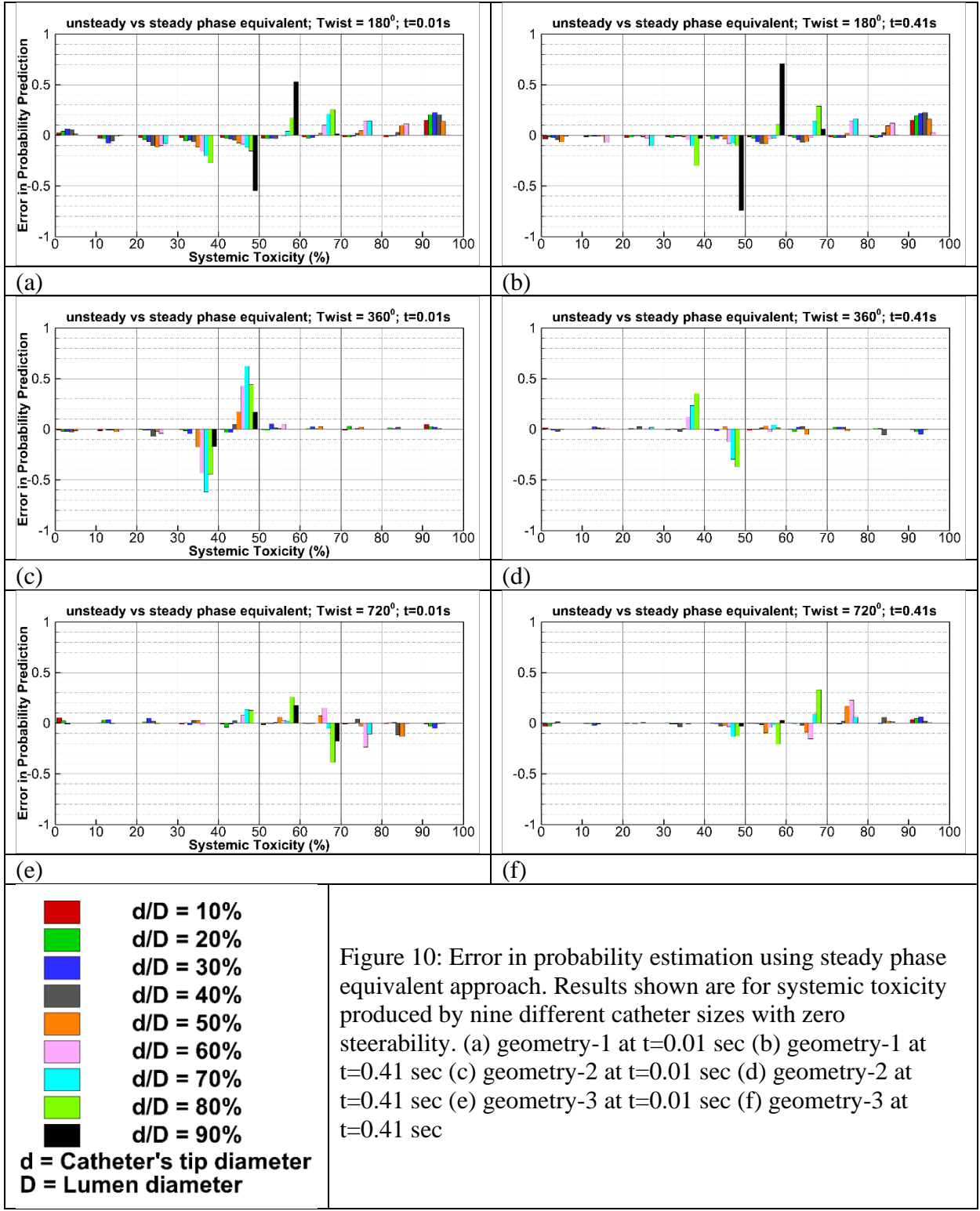

Figure 10: Error in probability estimation using steady phase equivalent approach. Results shown are for systemic toxicity produced by nine different catheter sizes with zero steerability. (a) geometry-1 at t=0.01 sec (b) geometry-1 at t=0.41 sec (c) geometry-2 at t=0.01 sec (d) geometry-2 at t=0.41 sec (e) geometry-3 at t=0.01 sec (f) geometry-3 at t=0.41 sec

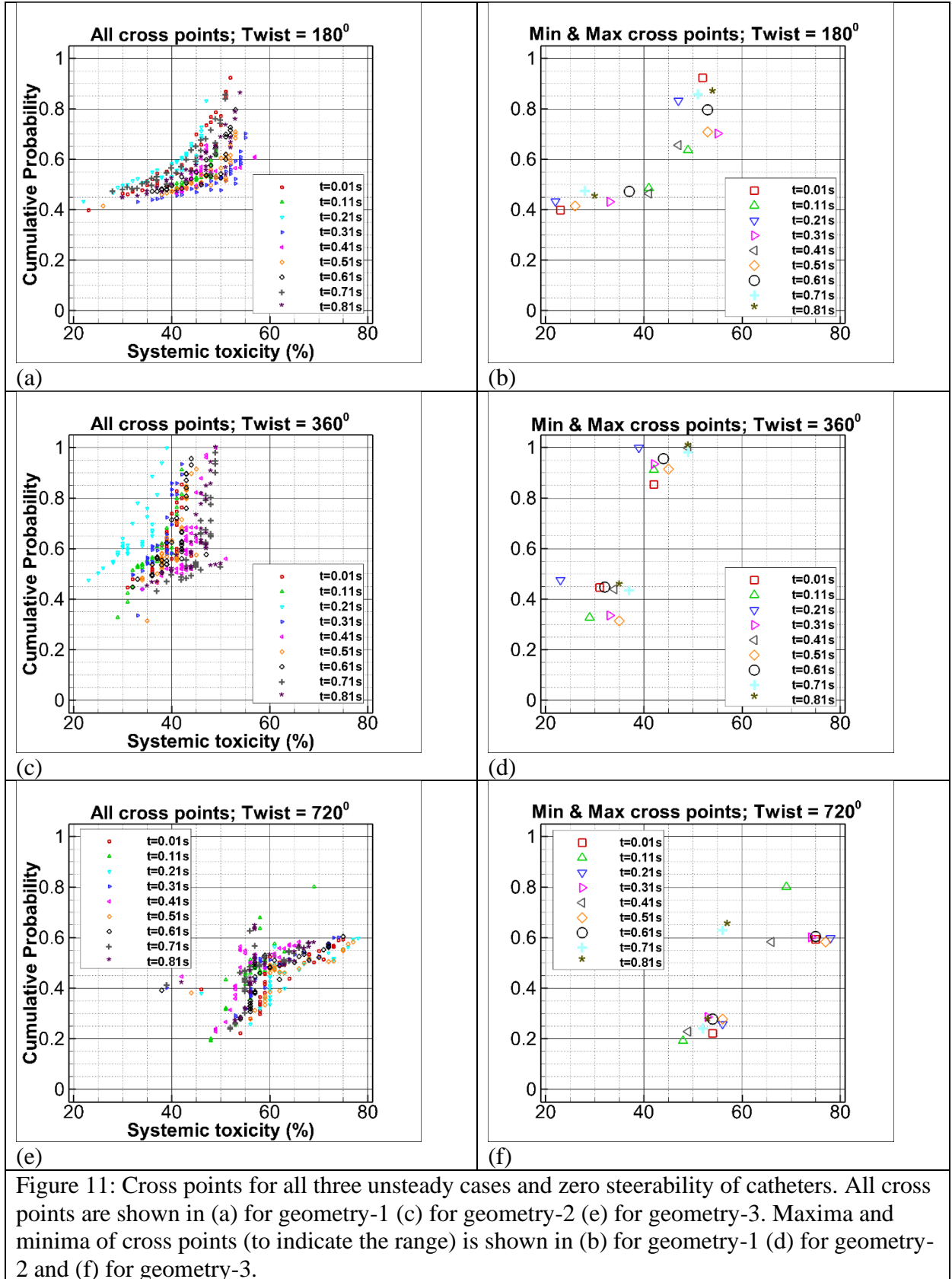

Figure 11: Cross points for all three unsteady cases and zero steerability of catheters. All cross points are shown in (a) for geometry-1 (c) for geometry-2 (e) for geometry-3. Maxima and minima of cross points (to indicate the range) is shown in (b) for geometry-1 (d) for geometry-2 and (f) for geometry-3.

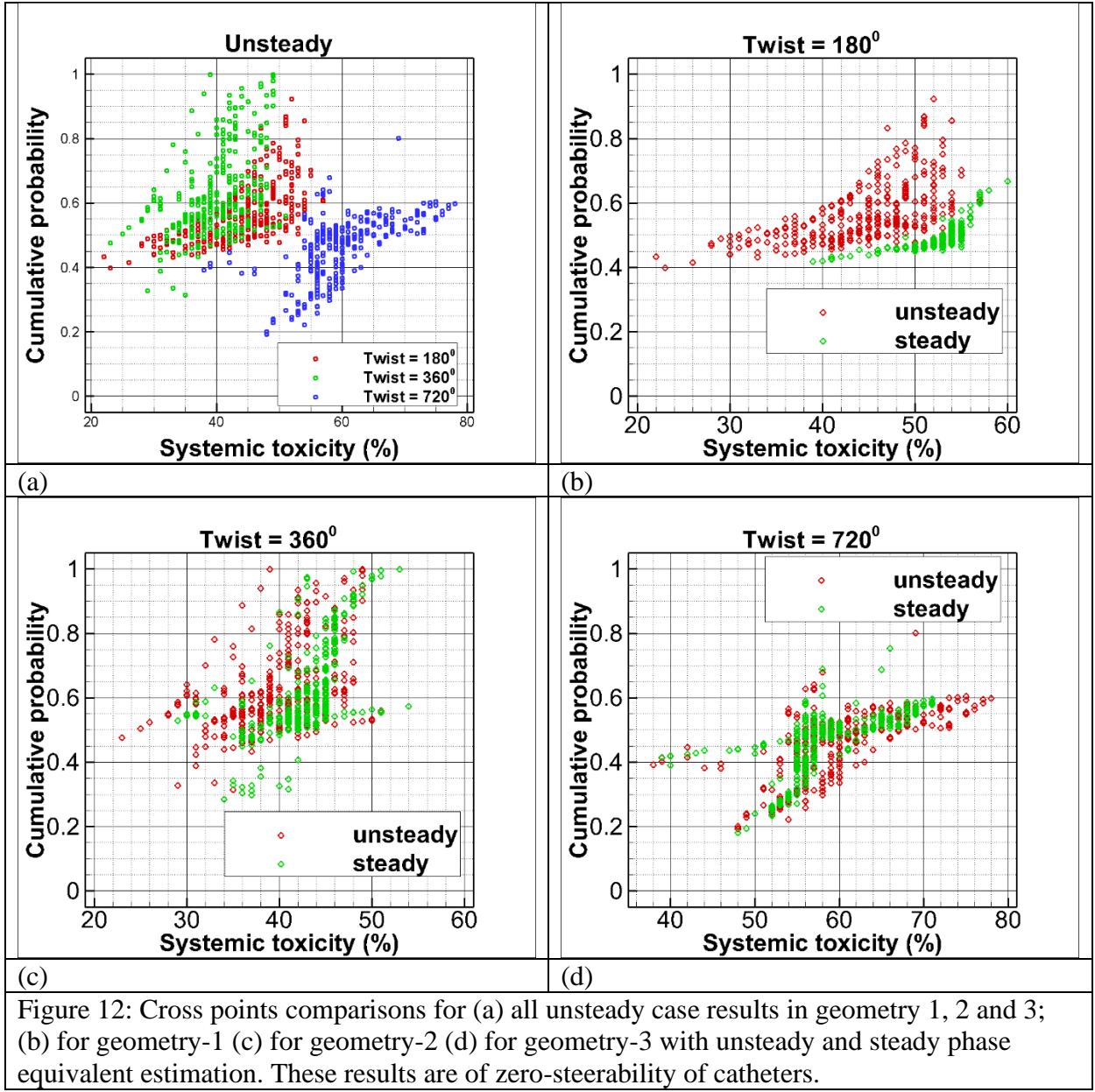

Figure 12: Cross points comparisons for (a) all unsteady case results in geometry 1, 2 and 3; (b) for geometry-1 (c) for geometry-2 (d) for geometry-3 with unsteady and steady phase equivalent estimation. These results are of zero-steerability of catheters.

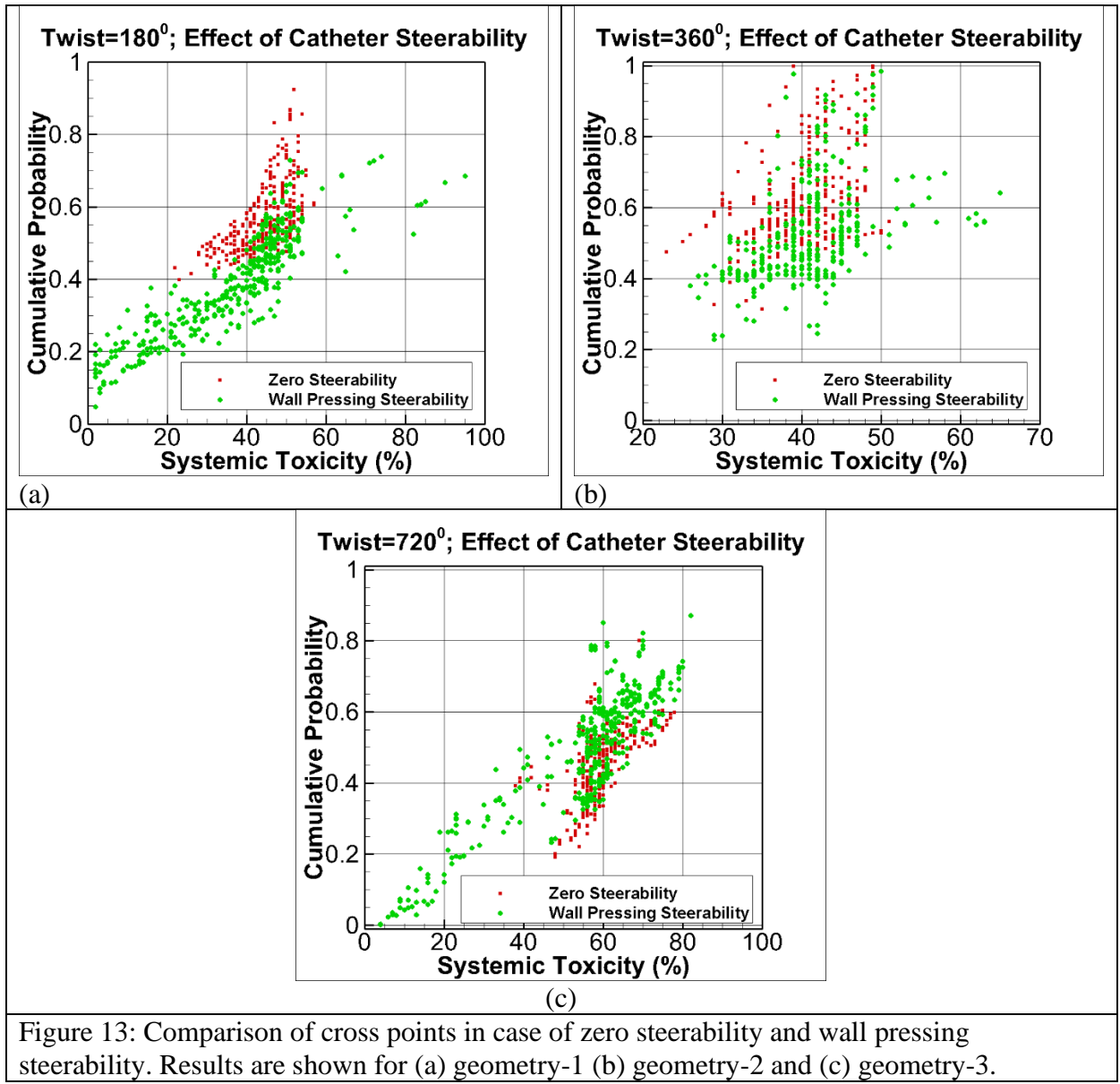

Figure 13: Comparison of cross points in case of zero steerability and wall pressing steerability. Results are shown for (a) geometry-1 (b) geometry-2 and (c) geometry-3.